\documentclass[11pt]{article}
\newcommand{\be}{\begin{equation}}
\newcommand{\ee}{\end{equation}}
\newcommand{\bea}{\begin{eqnarray}}
\newcommand{\eea}{\end{eqnarray}}
\newcommand{\sn}{{\rm sn}}

\newcommand{\dn}{{\rm dn}}
\newcommand{\cn}{{\rm cn}}
\newcommand{\sech}{{\rm sech}}

\begin{document}
\vspace{.5in}
\begin{center}
{\LARGE{\bf Domain Wall and Periodic Solutions of Coupled Asymmetric
Double Well Models}}
\end{center}

\vspace{.3in}
\begin{center}
{\LARGE{\bf Avinash Khare}} \\
{Institute of Physics, Bhubaneswar, Orissa 751005, India}
\end{center}

\begin{center}
{\LARGE{\bf Avadh Saxena}} \\
{Theoretical Division and Center for Nonlinear Studies, Los
Alamos National Laboratory, Los Alamos, NM 87545, USA}
\end{center}

\vspace{.9in}
{\bf {Abstract:}}

Coupled asymmetric double well ($a\phi^2-b\phi^3+c\phi^4$) one-dimensional
potentials arise in the context of first order phase transitions both
in condensed matter physics and field theory.  Here we provide an
exhaustive set of exact periodic  solutions of such a coupled asymmetric
model in terms of elliptic functions (domain wall arrays) and obtain
single domain wall solutions in specific limits.  We also calculate the
energy and interaction between solitons for various solutions.  Both
topological (kink-like at $T=T_c$) and nontopological (pulse-like for
$T\ne T_c$) domain wall solutions are obtained.  We relate some of these
solutions to domain walls in hydrogen bonded materials and also in the
field theory context. As a byproduct, we also obtain a new one parameter 
family of kink solutions of the uncoupled asymmetric double well model.

\newpage

\section{Introduction}

First order phase transitions are typically modeled by a triple well
($\phi^6$) free energy.  However, if a third order term becomes
symmetry allowed, one need not go to the sixth order term for the
transition to be of first order.  Instead an asymmetric double well
(or $\phi^2$-$\phi^3$-$\phi^4$) free energy is sufficient to drive
the transition (with or without an external field \cite{sanati1,sanati2}).
This situation occurs in body-centered cubic (bcc) to face-centered cubic
(fcc) reconstructive structural phase transitions in crystals, the 
$\omega$-phase transition in various elements and alloys \cite{sanati2}, 
isotropic to nematic phase transition in liquid crystals \cite{liquid} and 
in the hydrogen chains in hydrogen-bonded materials \cite{xu1,xu2,xu3,cheng}. 
 
Similar potentials arise in the field theory contexts as well with 
application to domain walls in carbon nanotubes and fullerenes \cite{field}.  
For triangular or hexagonal symmetry crystals two different order parameters 
(e.g. strain and shuffle) can couple with each other being described by an 
asymmetric double well \cite{jacobs,morse}.  We consider this situation in 
detail here and obtain several exact domain wall solutions of a coupled 
asymmetric double well model.  Note that in the uncoupled limit, the well 
known solutions are the (i) kink solution (when the condition $b_1^2=4a_1c_1$ 
is satisfied, see below) and (ii) the pulse solutions around the false 
vacuum both above and below $T_c$.  In this paper we generalize the well 
known kink solution of the uncoupled asymmetric double well model by 
obtaining a one parameter family of kink solutions of the same model.  We 
also obtain a few other solutions in the uncoupled limit, which to the best 
of our knowledge are not known in the literature. 

Lattice dynamical models of entropy driven first order transitions modeled 
by asymmetric double well potentials have been studied previously using 
both molecular dynamics and mean field theory \cite{kerr1,kerr2}.  Exactly 
solvable asymmetric double well potentials are also known \cite{selg}.  
There are many other physical contexts in which either the quantum 
mechanical eigenvalue problem or the domain wall solutions of the asymmetric 
double well potentials have been studied 
\cite{malomed,torii,cope,engle,theo,radosz,zhou}. Recently, we have obtained 
a large number of exact periodic domain wall solutions for the coupled 
$\phi^4$ \cite{cphi4} and the coupled $\phi^6$ \cite{cphi6} models.  Here 
we apply the same procedure to coupled asymmetric double well models. 
 
The paper is organized as follows.  In the next section (Sec. II) we present 
the coupled model and obtain the relevant equations of motion.  In Sec. III
we obtain several solutions including the kink lattice (``bright-bright'' 
soliton) solution at the transition temperature ($T=T_c$).  Section IV deals 
with the pulse lattice (``dark-dark") solutions both above and below $T_c$.  
We also obtain the energy of the soliton solutions and the asymptotic 
interaction between the solitons.  In Sec. V we consider a different variant 
of the model and obtain a few solutions of the corresponding coupled model.
Finally, in Sec. VI we conclude with remarks on physical relevance of these 
solutions and their comparison with the coupled $\phi^6$ and $\phi^4$ models.

\section{Model}

We consider a potential that is asymmetric in the two fields due to a 
linear-quadratic coupling in addition to a biquadratic coupling.
The potential we consider is given by ($c_1>0$, $c_2 \ge 0$)
\be\label{2.1}
V= (a_1 \phi^2 -b_1 \phi^3 +c_1\phi^4) + (a_2\psi^2-b_2\psi^3+c_2\psi^4) 
+d\phi \psi^2+e\phi^2 \psi^2\,,
\ee
where $a_{1,2},b_{1,2},c_{1,2},d,e$ are material or system dependent 
parameters.  The static field equations that follow from here are
\bea\label{2.2}
&&\phi_{xx}=2a_1\phi-3b_1 \phi^2+4c_1\phi^3+d\psi^2+2e \phi \psi^2 , 
\nonumber \\
&&\psi_{xx}=2a_2\psi-3b_2\psi^2+4c_2\psi^3+2d\phi \psi+2e\phi^2 \psi\,.
\eea
Observe that as long as $d \ne 0$, the two field equations are asymmetric and
hence bright-dark and dark-bright solitons would be distinct.

For the standard uncoupled model ($d=e=0$) one usually takes $a_1>0$ as it
then corresponds to a model for first order transition. 
The sign of $b_1$ merely decides if the other minimum is to the
right or left of $\phi=0$. Hence without any loss of generality,
one usually  restricts to $b_1>0$ and we will primarily stick to this choice
(i.e. $a_1,b_1,c_1 >0$) in this paper except in Sec. V.
In this case it is easy to show that while $\phi=0$ is the
only minimum in case $b_1^2 < (32/9)a_1c_1$, for $(32/9)a_1c_1 < b_1^2
<4a_1c_1$, $\phi=0$ is the absolute minimum while $\phi=\phi_c$ is the
local minimum, whereas for $b_1^2>4a_1c_1$, the opposite is true. Here
\be\label{2.3a}
\phi_c= \frac{[3b_1+\sqrt{(9b_1^2-32a_1c_1)}]}{8c_1}\,,
\ee
while the local maximum of the potential is always at $\phi_{m}$ and is 
given by
\be\label{2.4a}
\phi_m= \frac{[3b_1-\sqrt{(9b_1^2-32a_1c_1)}]}{8c_1}\, . 
\ee
At $b_1^2=4a_1c_1$ we have two degenerate minima at $\phi=0$ and 
$\phi=\phi_c$.
For the uncoupled model, 
it is well known that while at $T=T_c$ (i.e. $b_1^2=4a_1c_1$),
 one has a kink solution \cite{sanati1,sanati2}, for
$T> T_c$ as well as for $T<T_c$ 
one has a pulse solution around the local minimum \cite{sanati1,sanati2}, 
in a way related to the decay of the false vacuum.

It is perhaps not well appreciated that even for $a_1<0$,
the uncoupled model corresponds to a model for first order transition. In
particular, in that case while $\phi=0$ is always the local maximum,  
for $b_1>0$, $\phi_{c1}$ is the absolute minimum while $\phi_{c2}$ is the 
local minimum, while for $b_1 <0$, $\phi_{c2}$ and $\phi_{c1}$ interchange 
their roles.  Here $\phi_{c1,c2}$ are defined by
\be\label{2.5a}
\phi_{c1}= \frac{[3b_1+\sqrt{(9b_1^2+32a_1c_1)}]}{8c_1}\,,
\ee
\be\label{2.6}
\phi_{c2}= \frac{[3b_1-\sqrt{(9b_1^2+32a_1c_1)}]}{8c_1}\,.
\ee
At $b_1=0$ they are degenerate (the usual $\phi^4-\phi^2$ double well 
potential). 

In this paper, we primarily  focus on the case when $a_1 > 0$ and without any
loss of generality take $b_1>0$. Needless to say that from the consideration
of stability, $c_1$ is always taken to be positive throughout this paper.
However, in Sec. V we also write down a few solutions of the coupled model in
case $a_1 <0$.

We now write down the periodic soliton solutions of the coupled
continuum asymmetric $\phi^4$ models. We will also consider the solutions
in the (physically interesting) case of $b_2=c_2=0$. We will see that,
corresponding to the uncoupled limit of $b_1^2=4a_1c_1$ 
one has ten periodic soliton solutions in the coupled case which in
the hyperbolic limit correspond to three bright-bright,
two dark-dark and four dark-bright and bright-dark solutions, while one  
solution becomes a constant in this limit. In the physically interesting
case of $b_2=c_2=0$, only six out of these ten solutions survive.  On the 
other hand, for $T> (<)$ $T_c$, we are able to obtain only one solution of 
this coupled model which, depending on the value of the parameters, is 
related to the pulse solution around the false vacuum (local minimum), 
both above and below $T_c$.  It is interesting to note that some of the 
solutions exist only in the coupled model but not in the uncoupled limit.

For static solutions the energy is given by
\be\label{2.3}
E=\int \left[\frac{1}{2}\left(\frac{d\phi}{dx}\right)^2+\frac{1}{2}
\left(\frac{d\psi}{dx}\right)^2 +V(\phi,\psi)\right]\,dx \,,
\ee
where the limits of integration are from $-\infty$ to $\infty$ in the
case of hyperbolic solutions (i.e., single solitons) on the full line.
On the other hand, in the case of periodic solutions (i.e. soliton
lattices), the limits are from $-K(m)$ to $+K(m)$ or from $-2K(m)$ to
$+2K(m)$ depending on the period of the corresponding elliptic function.  
Here $K(m)$ [and
$E(m)$ below] denote the complete elliptic integral of the first (and
second) kind \cite{GR}. Using equations of motion, one can show that
for all of our solutions
\be\label{2.4}
V(\phi,\psi)=\left[\frac{1}{2}\left(\frac{d\phi}{dx}\right)^2
+\frac{1}{2}\left(\frac{d\psi}{dx}\right)^2\right] +C \,,
\ee
where the constant $C$ in general varies from solution to solution.
Hence the energy $\hat{E}=E-\int C\,dx$ is given by
\be\label{2.5}
\hat{E} \equiv E-\int C\,dx =\int\left[\left(\frac{d\phi}{dx}\right)^2
+\left(\frac{d\psi}{dx}\right)^2\right]\,dx .
\ee
Below we will give explicit expressions for energy in the case of several of 
the periodic solutions (and hence the corresponding hyperbolic
solutions). In each case we also provide an expression for the constant $C$.

\section{Solutions Corresponding to $b_1^2=4a_1c_1$ (in the Analogous 
Uncoupled Case)}

In this and the next section we consider the model with $a_1,b_1,c_1 >0$.
We look for the most general periodic solutions in terms of the Jacobi
elliptic functions sn($x,m$), cn($x,m$) and dn($x,m)$ \cite{GR} which in 
the decoupled limit correspond to $b_1^2=4a_1c_1$ (i.e. $T=T_c$).

{\bf Solution I}

It is
not difficult to show that
\be\label{3.1}
\phi =F+A\sn[D(x+x_0),m]\,,~~\psi =G+B\sn[D(x+x_0),m]\,,
\ee
is an exact solution provided the following eight coupled equations are
satisfied
\be
2a_1 F-3b_1F^2+4c_1 F^3+dG^2+2eFG^2=0\,,
\ee
\be
2a_1 A-6b_1AF+12c_1 F^2 A+2 BdG+2e AG^2
+4eFBG =-(1+m)AD^2\,,
\ee
\be
-3b_1A^2+12c_1 F A^2+dB^2+2e FB^2+4e ABG=0\,,
\ee
\be
2c_1 A^2+e B^2=mD^2\,,
\ee
\be
2a_2 G-3b_2G^2+4c_2 G^3+2dFG+2e GF^2 =0\,,
\ee
\be
2a_2 B-6b_2BG+12c_2 G^2 B+2dAG+2dFB+4e AFG+2e BF^2
=-(1+m)BD^2\,,
\ee
\be
-3b_2B^2+12c_2 G B^2+2dAB+2e GA^2+4e ABF
=0\,,
\ee
\be
2c_2 B^2+e A^2=mD^2\,.
\ee
Here $A$ and $B$ denote the amplitudes of the ``kink lattice", $F$
and $G$ are constants, $D$ is an inverse characteristic length and
$x_0$ is the (arbitrary) location of the kink.  Five of these equations
determine the five unknowns $A,B,D,F,G$ while the other three equations,
give three constraints between the eight parameters $a_{1,2},b_{1,2},
c_{1,2},d,e$.
In particular, $A$ and $B$ are given by
\be\label{3.2}
A^2=\frac{mD^2(2c_2-e)}{(4c_1 c_2 -e^2)}\,, ~~~
B^2=\frac{mD^2(2c_1-e)}{(4c_1 c_2 -e^2)}\,.
\ee

It may be noted here that in case both $F,G=0$ then no solution exists
so long as $a_1 > 0$.
However, a solution exists in case $G=0, F \ne 0$ or when $F=0, G\ne 0$,
both of which we now discuss one by one.

\noindent{\bf G=0, F$\ne$0:}

In this case $A,B$ are again given by Eq. (\ref{3.2}) while
$F$ and $D$ are given by
\bea\label{3.3}
&&F\equiv \phi_{m}=\frac{[3b_1-\sqrt{(9b_1^2-32a_1c_1)}]}{8c_1}\,, \nonumber 
\\
&&D^2=\frac{\sqrt{(9b_1^2-32a_1c_1)}\,
\bigg [3b_1-\sqrt{(9b_1^2-32a_1c_1)} \bigg ]}{8(1+m)c_1}\,.
\eea
Further there are three constraints given by
\bea\label{3.4}
&&\frac{2(d+2eF)}{3b_2}=\sqrt{\frac{(2c_1-e)}{(2c_2-e)}}\,,~~
(6b_1c_1-4c_1d-3eb_1)F=2(2a_2c_1+4a_1c_1-ea_1)\,, \nonumber \\
&&27b_2^2(b_1-4c_1F)=4(d+2eF)^2\,.
\eea
It is worth noting that the solution exists only when the value of $F$ 
corresponds to $\phi_{m}$ as given by Eq. (\ref{2.4a}). From one of the 
constraint [Eq.(\ref{3.4})] it follows that $b_1>4c_1F$, i.e. $b_1^2 > 4a_1c_1$.

\noindent{\bf Special case of $e^2 = 4c_1 c_2$}

One can show that the solution (\ref{3.1}) exists even in case $e^2=
4c_1 c_2$. It turns out that such a solution
exists only if
\be\label{3.6}
2c_1=2c_2=e\,.
\ee
In this case, $A$ and $B$ are determined from the relations
\be\label{3.7}
A^2+B^2=\frac{mD^2}{e}\,,~~3b_2B=2(d+2eF)A\,,
\ee
while $D$, $F$ are still given by Eq. (\ref{3.3}) and the two constraints
are
\be\label{3.4a}
27b_2^2(b_1-2eF)=4(d+2eF)^2\,,~~a_1+a_2+dF=0\,.
\ee

\noindent{\bf Interesting case of $b_2=c_2=0$}

Finally, let us discuss the physically interesting case
of $b_2=c_2=0$. In this case, solution (\ref{3.1}) with $G=0, F \ne 0$ 
exists only if $e>2c_1$. In
particular, in this case Eq. (\ref{3.1}) with $G=0$ is a solution to the
field Eq. (\ref{2.2}) provided
\be\label{3.7a}
A^2=\frac{ma_1}{(1+m)e}\,,~~B^2=\frac{(e-2c_1)ma_1}{(1+m)e^2}\,,~~
D^2=\frac{a_1}{(1+m)}\,,~~F=\sqrt{\frac{a_1}{4c_1}}\,,
\ee
and further if the following three constraints are satisfied
\be\label{3.8}
b_1^2=4a_1c_1\,,~~d^2=2e(a_1+2a_2)\,,~~b_1d+2a_1e=0\,.
\ee

\noindent{\bf F=0, G$\ne$0:}

Let us now discuss the solution in case $F=0$ but $G \ne 0$. In this case 
solution (\ref{3.1}) exists only if $d=0$ and $e <0$. We therefore 
write $e=-|e|$ and $A,B$ are now given by 
\be\label{3.2c}
A^2=\frac{mD^2(2c_2+|e|)}{(4c_1 c_2 -e^2)}\,, ~~~
B^2=\frac{mD^2(2c_1+|e|)}{(4c_1 c_2 -e^2)}\,.
\ee
while $G$ and $D$ are given by
\bea\label{3.3c}
&&G\equiv \psi_{m}=\frac{[3b_2-\sqrt{(9b_2^2-32a_2c_2)}]}{8c_2}\,, \nonumber 
\\
&&D^2=\frac{\sqrt{(9b_2^2-32a_2c_2)}\,
\bigg [3b_2-\sqrt{(9b_2^2-32a_2c_2)} \bigg ]}{8(1+m)c_2}\,.
\eea
Further there are three constraints given by
\bea\label{3.4c}
&&3b_1A=-4|e|BG\,,~~3B^2(b_2-4c_2G)=-2|e|GA^2 , \nonumber \\
&&3(2c_2+|e|)b_2G=2(2a_1c_2+4a_2c_2+|e|a_2)\,.
\eea
Note that one of the constraint implies that $b_2 \le 4c_2G$
which in turn implies that $b_2^2<4a_2 c_2$.

It is easily checked that {\it no} solution of the form (\ref{3.1}) exists 
when either  $e^2=4c_1c_2$ or if $b2=c_2=0$.

\noindent{\bf m=1}

In the limiting case $G=0, F \ne 0, m=1$, 
we have a bright-bright soliton solution given by
\be\label{3.5}
\phi =F+A\tanh[D(x+x_0)]\,,~~\psi =B\tanh[D(x+x_0)]\,,
\ee
with $A$, $B$, $F$ and $D$ given by Eqs. (\ref{3.2}) and (\ref{3.3}) with
$m=1$
while the three constraints are again given by Eq. (\ref{3.4}).
A similar solution also exists in case $F=0, G \ne 0$.

\noindent{\bf Uncoupled Case}

For completeness, it may be worthwhile to write down the solution of the
uncoupled field Eq. (\ref{2.2}), i.e. when $d=e=0$. It is easily shown that
in this case the solution (\ref{3.1}) reduces to \cite{sanati1,sanati2} 
\be\label{3.9}
\phi= \sqrt{\frac{a_1}{4c_1}} \bigg (1+\frac{2m}{(1+m)}
\sn\left[\sqrt{\frac{a_1}{(1+m)}}(x+x_0),m\right] \bigg )\,,
\ee
and one is at the transition temperature $T=T_c$ since the parameters 
$a_1,b_1,c_1$ satisfy $b_1^2=4a_1c_1$. At $m=1$, this reduces to the well 
known kink solution 
\cite{sanati1,sanati2}  
\be\label{3.10}
\phi= \sqrt{\frac{a_1}{4c_1}}  \bigg [1+
\tanh(\sqrt{\frac{a_1}{2}}[x+x_0]) \bigg ]\,.
\ee

\noindent{\bf Energy:} Corresponding to the periodic solution
(\ref{3.1}) with $G=0$ 
the energy is same irrespective of whether $F=0, G \ne 0$ 
or $G=0, F \ne 0$ or if both of them are zero. Only the value of $C$ 
is different for the different cases. For example, for $G=0, F \ne 0$ 
%\be
%\phi =F+A\sn[D(x+x_0),m]\,,~~\psi =B\sn[D(x+x_0),m]\,,
%\ee
the energy $\hat{E}$ and the constant $C$ are given by
\bea\label{3.11}
&&\hat{E}=\frac{2(A^2+B^2)D}{3m}[(1+m)E(m)-(1-m)K(m)]\,, \nonumber \\
&&C=F^2[a_1-b_1F+c_1F^2]-\frac{1}{2}(A^2+B^2)D^2\,.
\eea
On the other hand, in case $F=0, G \ne 0$ then $C$ is given by
\be\label{3.11c}
C=G^2[a_2-b_2G+c_2G^2]-\frac{1}{2}(A^2+B^2)D^2\,.
\ee

On using the expansion formulas for $E(m)$ and $K(m)$ around $m=1$ as
given in \cite{GR}
\be\label{3.13}
K(m)=\ln\left(\frac{4}{\sqrt{1-m}}\right)+\frac{(1-m)}{4}\left[\ln\left(
\frac{4}{\sqrt{1-m}}\right)-1\right]+...\,,
\ee
\be\label{3.14}
E(m)=1+\frac{(1-m)}{2}\left[\ln\left(\frac{4}{\sqrt{1-m}}\right)-\frac{1}{2}
\right]+...\,,
\ee
for $m$ near one, the energy of the periodic solution can be rewritten
as the energy of the corresponding hyperbolic (bright-bright) soliton
solution [Eq. (\ref{3.5})]
%\be
%\phi =F+A\tanh[D(x+x_0)]\,,~~\psi =B\tanh[D(x+x_0)]\,,
%\ee
plus the interaction energy. We find
\be\label{3.15}
\hat{E}=E_{kink}+E_{int}=(A^2+B^2)D
\left[\frac{4}{3}+\frac{(1-m)}{3}\right]\,.
\ee
The interaction energy
vanishes at exactly $m=1$, as it should.

{\bf Solution II}

A different type of solution (i.e. a pulse lattice) given by
\be\label{3.16}
\phi =F+A\cn[D(x+x_0),m]\,,~~\psi =G+B\cn[D(x+x_0),m]\,,
\ee
is an exact solution provided the following eight coupled equations are
satisfied
\be
2a_1 F-3b_1F^2+4c_1 F^3+dG^2+2eFG^2=0\,,
\ee
\be
2a_1 A-6b_1AF+12c_1 F^2 A+2 BdG+2e AG^2
+4eFBG =(2m-1)AD^2\,,
\ee
\be
-3b_1A^2+12c_1 F A^2+dB^2+2e FB^2+4e ABG=0\,,
\ee
\be
2c_1 A^2+e B^2=-mD^2\,,
\ee
\be
2a_2 G-3b_2G^2+4c_2 G^3+2dFG+2e GF^2 =0\,,
\ee
\be
2a_2 B-6b_2BG+12c_2 G^2 B+2dAG+2dFB+4e AFG+2e BF^2
=(2m-1)BD^2\,,
\ee
\be
-3b_2B^2+12c_2 G B^2+2dAB+2e GA^2+4e ABF
=0\,,
\ee
\be
2c_2 B^2+e A^2=-mD^2\,.
\ee
Notice that two of these equations are meaningful only if $e < 0$
since from stability considerations, $c_1>0, c_2 \ge 0$. Hence, we write 
$e=-|e|$.

Five of these equations
determine the five unknowns $A,B,D,F,G$ while the other three equations,
give three constraints between the eight parameters $a_{1,2},b_{1,2},
c_{1,2},d,e$.
In particular, $A$ and $B$ are given by
\be\label{3.17}
A^2=\frac{mD^2(2c_2+|e|)}{(e^2-4c_1 c_2)}\,, ~~~
B^2=\frac{mD^2(2c_1+|e|)}{(e^2-4c_1 c_2)}\,.
\ee

Unlike the previous case, it turns out that in this case a solution exists
both when $F=G=0$, $G=0$ or when $F \ne 0$ or if $F=0, G \ne 0$ 
and we discuss all three cases one by one.

\noindent{\bf F=G=0}

In this case $A,B$ are again given by Eq. (\ref{3.17}) while $D$, the 
characteristic inverse length, is given by
\be\label{3.18}
D^2=\frac{2a_1}{(2m-1)}\,,
\ee
while the three constraints are
\be\label{3.19}
a_1=a_2\,,~~4d^3=27b_1b_2^2\,,~~
\frac{3b_1}{d}=\sqrt{\frac{(2c_1+|e|)}{(2c_2+|e|)}}\,.
\ee
Since $a_1 >0$, such a solution exists only if $m>1/2$.
It is easily checked that such a solution (with $F=G=0$) does 
not exist in case either $b_2=c_2=0$ or if $e^2=4c_1c_2$.

What happens if $m=1/2$? It is easily checked that in that case (\ref{3.16})
with $F=G=0$ is still a solution provided $a_1=a_2=0$, $D$ is undetermined,
$A,B$ are given by Eq. (\ref{3.17}) and the two remaining constraints are 
given by Eq. (\ref{3.19}).

\noindent{\bf G=0, F$\ne$0:}

In this case (\ref{3.16}) is a solution with 
$A,B$ again given by Eq. (\ref{3.17}) while $F$ and $D$ are given by
\bea\label{3.20}
&&F \equiv \phi_c =\frac{[3b_1+\sqrt{(9b_1^2-32a_1c_1)}]}{8c_1}\,, \nonumber 
\\
&&D^2=\frac{\sqrt{(9b_1^2-32a_1c_1)}\,
[3b_1+\sqrt{(9b_1^2-32a_1c_1)}]}{8(2m-1)c_1}\,.
\eea
Further there are three constraints given by
\bea\label{3.21}
&&\frac{(2|e|F-d)}{3(4c_1F-b_1)}=\frac{(2c_1+|e|)}{(2c_2+|e|)}\,,~~
(6b_1c_1-4c_1d+3|e|b_1)F=2(2a_2c_1+4a_1c_1+|e|a_1)\,, \nonumber \\
&&27b_2^2(4c_1F-b_1)=4(2|e|F-d)^2\,.
\eea

What happens if $m=1/2$? It is easily checked that in that case (\ref{3.16})
with $G=0$ is still a solution provided 
$A,B$ are given by Eq. (\ref{3.17}), $D$ is undetermined  while the  
constraints are
\be\label{3.21a}
3b_1F=4a_1\,,~~9b_1^2=32a_1c_1\,,~~
3b_2B=2dA\,,~~(2|e|F-d)B^2=4c_1FA^2\,,~~a_2+dF=|e|F^2\,.
\ee

Interestingly enough, there is also a solution in case $m<1/2$. In particular,
it is easily checked that (\ref{3.16}) is a solution with 
$A,B$ again given by Eq. (\ref{3.17}) while $F$ and $D$ are now given by
\bea\label{3.20a}
&&F \equiv \phi_m =\frac{[3b_1-\sqrt{(9b_1^2-32a_1c_1)}]}{8c_1}\,, \nonumber 
\\
&&D^2=\frac{\sqrt{(9b_1^2-32a_1c_1)}\,
[3b_1-\sqrt{(9b_1^2-32a_1c_1)}]}{8(1-2m)c_1}\,.
\eea
Further the three constraints are again given by Eq. (\ref{3.21}).

{\bf Interesting case of $b_2=c_2=0$}

Finally, let us discuss the physically interesting case
of $b_2=c_2=0$.
It is easy to see that now a solution of the form (\ref{3.16})
with $G=0$ is possible only if $m <1/2$. In particular, such a solution
exists if
\be\label{3.22a}
A^2=\frac{mD^2}{|e|}\,, ~~~
B^2=\frac{mD^2(2c_1+|e|)}{e^2}\,,~~ D^2=\frac{a_1}{(1-2m)}\,,
\ee
and if further the following constraints are satisfied
\be\label{3.22b}
b_1^2=4a_1c_1\,,~~db_1=2|e|a_1\,,~~d^2=-2|e|(a_1+2a_2)\,.
\ee
Thus with $a_1>0$, a solution of the form (\ref{3.16}) 
with $b_2=c_2=0$ can exist only if
$m<1/2$ and  $a_2<0$.

It is easily checked that no solution is however 
possible in the special case of $4c_1c_2=e^2$.

\noindent{\bf F=0, G$\ne$0:}

In this case (\ref{3.16}) is a solution only if $d=0$ with 
$A,B$ again given by Eq. (\ref{3.17}) while $G$ and $D$ are given by
\bea\label{3.20c}
&&G \equiv \psi_c =\frac{[3b_2+\sqrt{(9b_2^2-32a_2c_2)}]}{8c_2}\,, \nonumber 
\\
&&D^2=\frac{\sqrt{(9b_2^2-32a_2c_2)}\,
[3b_2+\sqrt{(9b_2^2-32a_2c_2)}]}{8(2m-1)c_2}\,.
\eea
Further there are three constraints given by
\be\label{3.21c}
3b_1A=-4|e|BG\,,~~(4c_2G-b_2)B^2=2|e|GA^2\,,~~2a_1-2|e|G^2=(2m-1)D^2\,.
\ee
Note that this solution exists only if $m>1/2$. 

What happens if $m=1/2$? It is easily checked that in that case (\ref{3.16})
with $F=0$ is still a solution provided 
$A,B$ are given by Eq. (\ref{3.17}), $D$ is undetermined  and the 
constraints are
\be\label{3.21d}
3b_1A=-4|e|BG\,,~~a_1=|e|G^2\,,~~b_2B^2=4|e|GA^2\,,
~~3b_2G=4a_2\,,~~ 9b_2^2=32a_2c_2\,. 
\ee

Interestingly enough, there is also a solution in case $m<1/2$. In particular,
it is easily checked that (\ref{3.16}) is a solution with 
$A,B$ again given by Eq. (\ref{3.17}) while $G$ and $D$ are now given by
\bea\label{3.21e}
&&G \equiv \psi_m =\frac{[3b_2-\sqrt{(9b_2^2-32a_2c_2)}]}{8c_2}\,, \nonumber 
\\
&&D^2=\frac{\sqrt{(9b_2^2-32a_2c_2)}\,
[3b_2-\sqrt{(9b_2^2-32a_2c_2)}]}{8(1-2m)c_2}\,.
\eea
The constraints are again given by Eq. (\ref{3.21c}). Note that one of the
constraint can only be satisfied if $b_2^2 < 4a_2c_2$.
Even in this case, no solution of the form (\ref{3.16}) with $F=0, G \ne 0$
exists either when $e^2=4c_1c_2$ or when $b_2=c_2=0$.

\noindent{\bf m=1}

In this limiting case we have a dark-dark soliton solution given by
\be\label{3.22}
\phi =F+A\sech[D(x+x_0)]\,,~~\psi =G+B\sech[D(x+x_0)]\,,
\ee
where $F$ and/or $G$ 
may or may not be zero depending on the solution that we are considering.

It is worth emphasizing that while solution (\ref{3.16}) exists in the 
coupled case, there is {\it no} analogous solution to the corresponding 
uncoupled model, i.e. no solution is possible in the case of the uncoupled 
field Eq. (\ref{2.2}) i.e. when $d=e=0$, so long as $c_1>0$.

\noindent{\bf Energy:} Corresponding to the periodic solution
(\ref{3.16}) the energy is same but only the value of $C$ changes depending
on if $F=G=0$ or $G=0, F \ne 0$ or $F=0, G \ne 0$. For example, if 
$G=0, F \ne 0$ then 
%\be
%\phi =F+A\cn[D(x+x_0),m]\,,~~\psi =B\cn[D(x+x_0),m]\,,
%\ee
the energy $\hat{E}$ and the constant $C$ are given by
\bea\label{3.23}
&&\hat{E}=\frac{2(A^2+B^2)D}{3m}[(2m-1)E(m)+(1-m)K(m)]\,, \nonumber \\
&&C=F^2[a_1-b_1F+c_1F^2]-\frac{1}{2}(1-m)(A^2+B^2)D^2\,.
\eea
On the other hand, if $F=0, G \ne 0$ then $C$ is given by
\be\label{3.23c}
C=G^2[a_2-b_2G+c_2G^2]-\frac{1}{2}(1-m)(A^2+B^2)D^2\,,
\ee
while if $F=G=0$ then $C$ is simply given by
\be\label{3.23d}
C=-\frac{1}{2}(1-m)(A^2+B^2)D^2\,. 
\ee

On using the expansion formulas for $E(m)$ and $K(m)$ around $m=1$ as
given by Eqs. (\ref{3.13}) and (\ref{3.14}),
for $m$ near one, the energy of the periodic solution can be rewritten
as the energy of the corresponding hyperbolic (dark-dark) soliton
solution [Eq. (\ref{3.22})]
%\be
%\phi =F+A\tanh[D(x+x_0)]\,,~~\psi =B\tanh[D(x+x_0)]\,,
%\ee
plus the interaction energy. We find
\be\label{3.24}
\hat{E}=E_{pulse}+E_{int}=(A^2+B^2)D \left[\frac{2}{3}
-\frac{5(1-m)}{6}+(1-m)\ln (\frac{4}{\sqrt{1-m}}) \right]\,.
\ee
Note that this solution exists only when $e<0$ and $e^2>4c_1c_2$.
The interaction energy
vanishes at exactly $m=1$, as it should.

{\bf Solution III}

Yet another pulse lattice solution given by
\be\label{3.25}
\phi =F+A\dn[D(x+x_0),m]\,,~~\psi =G+B\dn[D(x+x_0),m]\,,
\ee
is an exact solution provided the following eight coupled equations are
satisfied
\be
2a_1 F-3b_1F^2+4c_1 F^3+dG^2+2eFG^2=0\,,
\ee
\be
2a_1 A-6b_1AF+12c_1 F^2 A+2 BdG+2e AG^2
+4eFBG =(2-m)AD^2\,,
\ee
\be
-3b_1A^2+12c_1 F A^2+dB^2+2e FB^2+4e ABG=0\,,
\ee
\be
2c_1 A^2+e B^2=-D^2\,,
\ee
\be
2a_2 G-3b_2G^2+4c_2 G^3+2dFG+2e GF^2 =0\,,
\ee
\be
2a_2 B-6b_2BG+12c_2 G^2 B+2dAG+2dFB+4e AFG+2e BF^2
=(2-m)BD^2\,,
\ee
\be
-3b_2B^2+12c_2 G B^2+2dAB+2e GA^2+4e ABF
=0\,,
\ee
\be
2c_2 B^2+e A^2=-D^2\,.
\ee
Notice that two of these equations are meaningful only if $e<0$, since
from stability considerations, $c_1>0, c_2 \ge 0$. Hence we write
$e=-|e|$.

Five of these equations
determine the five unknowns $A,B,D,F,G$ while the other three equations,
give three constraints between the eight parameters $a_{1,2},b_{1,2},
c_{1,2},d,e$.
In particular, $A$ and $B$ are given by
\be\label{3.26}
A^2=\frac{D^2(2c_2+|e|)}{(e^2-4c_1 c_2)}\,, ~~~
B^2=\frac{D^2(2c_1+|e|)}{(e^2-4c_1 c_2)}\,.
\ee

Akin to the solution II, in this case too, a solution exists
when $F=G=0$ or  $G=0, F \ne 0$ or when $F=0, G \ne 0$ 
and we discuss all the cases one by one.

\noindent{\bf F=G=0}

In this case $A,B$ are again given by Eq. (\ref{3.26}) while $D$ is given by
\be\label{3.27}
D^2=\frac{2a_1}{(2-m)}\,,
\ee
and the three constraints are
\be\label{3.28}
a_1=a_2\,,~~4d^3=27b_1b_2^2\,,~~\frac{3b_1}{d}=
\sqrt{\frac{(2c_1+|e|)}{(2c_2+|e|)}}\,.
\ee

It may be noted that such a solution with $F=G=0$ will not exist in case 
either
$b_2=c_2=0$ or if $e^2=4c_1c_2$.

\noindent{\bf G=0, F$\ne$0:}

In this case $A,B$ are again given by Eq. (\ref{3.26}) while
$F$ and $D$ are given by
\bea\label{3.29}
&&F\equiv \phi_c =\frac{3b_1+\sqrt{(9b_1^2-32a_1c_1)}}{8c_1}\,, \nonumber \\
&&D^2=\frac{\sqrt{(9b_1^2-32a_1c_1)}\,
[3b_1+\sqrt{(9b_1^2-32a_1c_1)}]}{8(2-m)c_1}\,.
\eea
Further there are three constraints given by
\bea\label{3.30}
&&\frac{(2|e|F-d)}{3(4c_1F-b_1)}=\frac{(2c_1+|e|)}{(2c_2+|e|)}\,,~~
(6b_1c_1-4c_1d+3|e|b_1)F=2(2a_2c_1+4a_1c_1+|e|a_1)\,, \nonumber \\
&&27b_2^2(4c_1F-b_1)=4(2|e|F-d)^2\,.
\eea
It is worth noting that since $b_1-4c_1F<0$, hence $F>d/2|e|$. It is easily 
checked that no solution is however possible in case $b_2=c_2=0$ or if 
$4c_1c_2=e^2$.

\noindent{\bf m=1}

In this limiting case we have the same dark-dark soliton solution
as given by Eq. (\ref{3.22}).

\noindent{\bf F=0, G$\ne$0:}

Solution (\ref{3.25}) with $F=0, G \ne 0$ is only possible if $d=0$.
In this case $A,B$ are again given by Eq. (\ref{3.26}) while
$G$ and $D$ are given by
\bea\label{3.29c}
&&G\equiv \psi_c =\frac{3b_2+\sqrt{(9b_2^2-32a_2c_2)}}{8c_2}\,, \nonumber \\
&&D^2=\frac{\sqrt{(9b_2^2-32a_2c_2)}\,
[3b_2+\sqrt{(9b_2^2-32a_2c_2)}]}{8(2-m)c_2}\,.
\eea
Further there are three constraints given by
\be\label{3.30c}
3b_1A=4|e|BG\,,~~2a_1-2|e|G^2=(2-m)D^2\,,~~3(4c_2G-b_2)B^2=2|e|GA^2\,.
\ee
Again, it is easily checked that no solution is however 
possible in case $b_2=c_2=0$ or if $4c_1c_2=e^2$.

It may be noted that, as in the previous case, while there exists a solution 
(\ref{3.25}) in the coupled case, no such solution exists in the corresponding
uncoupled case, i.e. no solution exists in the case of
the uncoupled field Eq. (\ref{2.2}) i.e. when $d=e=0$, so long as $c_1>0$.

\noindent{\bf Energy:} Corresponding to the periodic solution
(\ref{3.25}), the energy is same but only $C$ is different depending
on if $F=G=0$, or $G=0, F \ne 0$ or $F=0, G \ne 0$. For example,  
with $G=0, F \ne 0$,
%\be
%\phi =F+A\cn[D(x+x_0),m]\,,~~\psi =B\cn[D(x+x_0),m]\,,
%\ee
the energy $\hat{E}$ and the constant $C$ are given by
\bea\label{3.31}
&&\hat{E}=\frac{2(A^2+B^2)D}{3}[(2-m)E(m)-(1-m)K(m)]\,, \nonumber \\
&&C=F^2[a_1-b_1F+c_1F^2]+\frac{1}{2}(1-m)(A^2+B^2)D^2\,.
\eea
On the other hand, if $F=0, G \ne 0$, then $C$ is given by
\be
C=G^2[a_2-b_2G+c_2G^2]+\frac{1}{2}(1-m)(A^2+B^2)D^2\,,
\ee
while if $F=G=0$ then $C$ is simply given by
\be
C=\frac{1}{2}(1-m)(A^2+B^2)D^2\,,
\ee

On using the expansion formulas for $E(m)$ and $K(m)$ around $m=1$ as
given by Eqs. (\ref{3.13}) and (\ref{3.14}),
for $m$ near one, the energy of the periodic solution can be rewritten
as the energy of the corresponding hyperbolic (dark-dark) soliton
solution [Eq. (\ref{3.22})]
%\be
%\phi =F+A\tanh[D(x+x_0)]\,,~~\psi =B\tanh[D(x+x_0)]\,,
%\ee
plus the interaction energy. We find
\be\label{3.32}
\hat{E}=E_{pulse}+E_{int}=(A^2+B^2)D \left[\frac{2}{3}
-\frac{(1-m)}{2}-(1-m)\ln\left(\frac{4}{\sqrt{1-m}}\right) \right]\,.
\ee
Note that this solution exists only when $e<0, e^2>4c_1c_2$.
The interaction energy vanishes at exactly $m=1$, as it should.
Unlike the dark-dark $\cn-\cn$ and $\dn-\dn$ solutions, it turns out that
the (mixed) dark-dark periodic soliton solutions of the form $\dn-\cn$ and
$\cn-\dn$, do not exist.

\noindent{\bf Solution IV}

We now discuss two bright-dark and two dark-bright periodic soliton
solutions.
We will see that for these four solutions
$G$ is necessarily
zero while $F$ is necessarily nonzero. Let us discuss these solutions one by
one. In particular, it is easily shown that
\be\label{3.33}
\phi =F+A\sn[D(x+x_0),m]\,,~~\psi =G+B\cn[D(x+x_0),m]\,,
\ee
is an exact solution provided
\be\label{3.34}
G=0\,,~~b_2=0\,,~~ d+2eF=0\,,
\ee
 and the following six
 coupled equations are
satisfied
\be
2a_1 F-3b_1F^2+4c_1 F^3=0\,,
\ee
\be
2a_1 -6b_1F+12c_1 F^2 +2eB^2
=-(1+m)D^2\,,
\ee
\be
-3b_1A^2+12c_1 F A^2=0\,,
\ee
\be
2c_1 A^2-e B^2=mD^2\,,
\ee
\be
2a_2 +2dF+2e(A^2+F^2)
=(2m-1)D^2\,,
\ee
\be
2c_2 B^2-e A^2=-mD^2\,.
\ee
Three of these equations
determine the three unknowns $A,B,D$ while the other three equations,
give three constraints between the seven parameters $a_{1,2},b_{1},
c_{1,2},d,e$.
In particular, $A$ and $B$ are given by
\be\label{3.35}
A^2=\frac{mD^2(e-2c_2)}{(e^2-4c_1 c_2)}\,, ~~~
B^2=\frac{mD^2(2c_1-e)}{(e^2-4c_1 c_2)}\,,
\ee
while the inverse characteristic length $D$ is given by
\be\label{3.36}
D^2=\frac{(e^2-4c_1c_2)a_1}{[(1+m)(e^2-4c_1c_2)+2em(2c_1-e)]}\,.
\ee
Further, the three constraints are
\be\label{3.37}
b_1^2=4a_1c_1\,,~~2a_1e+db_1=0\,,~~(2m-1)D^2=2a_2+2eA^2+dF\,.
\ee

\noindent{\bf m=1}

In this limiting case we have a bright-dark soliton solution given by
\be\label{3.38}
\phi =F+A\tanh[D(x+x_0)]\,,~~\psi =B\sech[D(x+x_0)]\,,
\ee
with $A$, $B$ and $D$ given by Eqs. (\ref{3.35}) and (\ref{3.36})
with $m=1$
while the three constraints are again given by Eq. (\ref{3.37}).

\noindent{\bf Special case of $e^2 = 4c_1 c_2$}

One can show that the solution (\ref{3.33}) exists even in case $e^2=
4c_1 c_2$. It turns out  such a solution exists only if Eqs. (\ref{3.6}) 
and (\ref{3.34}) are satisfied.  In this case, $A$, $B$ and $D$ are given by
\be\label{3.40}
A^2= \frac{[ma_1-2(1-m)a_2]}{2em}\,,~~
B^2= \frac{[ma_1-2(1+m)a_2]}{2em}\,,~~
D^2=\frac{2a_2}{m}\,,
\ee
while the constraints are
\be\label{3.41}
d=-b_1\,,~~b_1^2=4a_1c_1\,.
\ee

\noindent{\bf Interesting case of $b_2=c_2=0$}

Finally, let us discuss the physically interesting case
of $b_2=c_2=0$. 
In this case Eq. (\ref{3.33}) with $G=0$ is a solution to the
field Eqs. (\ref{2.2}) provided
\be\label{3.42}
A^2=\frac{m(d^2-4ea_2)}{2e^2}\,,~~B^2=\frac{(2c_1-e)(d^2-4ea_2)}{2e^3}\,,~~
D^2=\frac{(d^2-4ea_2)}{2e}\,,~~F=-\frac{d}{2e}\,,
\ee
and further if the following three constraints are satisfied
\be\label{3.43}
b_1^2=4a_1c_1\,,~~(d^2-4ea_2)[4mc_1+(1-m)e]=2a_1 e^2\,,~~b_1d+2a_1e=0\,.
\ee

\noindent{\bf Energy:} Corresponding to the periodic solution
(\ref{3.33})
%\be
%\phi =F+A\cn[D(x+x_0),m]\,,~~\psi =B\cn[D(x+x_0),m]\,,
%\ee
the energy $\hat{E}$ and the constant $C$ are given by
\bea\label{3.44}
&&\hat{E}=\frac{2D}{3m}\bigg ([(2-m)mB^2+(1+m)A^2]E(m)
-(1-m)(A^2+2B^2)K(m) \bigg )\,, \nonumber \\
&&C=F^2[a_1-b_1F+c_1F^2]-\frac{1}{2} A^2 D^2 +B^2[a_2+dF+eF^2+c_2 B^2]\,.
\eea

On using the expansion formulas for $E(m)$ and $K(m)$ around $m=1$ as
given by Eqs. (\ref{3.13}) and (\ref{3.14}),
for $m$ near one, the energy of the periodic solution can be rewritten
as the energy of the corresponding hyperbolic (bright-dark) soliton
solution [Eq. (\ref{3.38})]
%\be
%\phi =F+A\tanh[D(x+x_0)]\,,~~\psi =B\tanh[D(x+x_0)]\,,
%\ee
plus the interaction energy. We find
\be\label{3.45}
\hat{E}=E_{soliton}+E_{int}=D \left[\frac{2}{3}(2A^2+B^2)
+\frac{(1-m)}{6}(2A^2+3B^2)-(1-m)B^2\ln\left(\frac{4}{\sqrt{1-m}}\right)\right]\,.
\ee
Note that this solution exists when either (i) $2c_1\ge e \ge 2c_2$ and
$e^2 \ge 4c_1c_2$ or if (ii) $2c_2 \ge e \ge 2c_1$ and $4c_1c_2 \ge e^2$
or if (iii) $c_2=0, 2c_1 > e$.
The interaction energy
vanishes at exactly $m=1$, as it should.

\noindent{\bf Solution V}

It is easy to show that another such (kink- and pulse-like) solution is
\be\label{3.46}
\phi =F+A\sn[D(x+x_0),m]\,,~~\psi =G+B\dn[D(x+x_0),m]\,.
\ee
This is an exact solution provided
\be\label{3.47}
G=0\,,~~b_2=0\,,~~ d+2eF=0\,,
\ee
 and the following six
 coupled equations are
satisfied
\be
2a_1 F-3b_1F^2+4c_1 F^3=0\,,
\ee
\be
2a_1 -6b_1F+12c_1 F^2 +2eB^2
=-(1+m)D^2\,,
\ee
\be
-3b_1A^2+12c_1 F A^2=0\,,
\ee
\be
2c_1 A^2-e mB^2=mD^2\,,
\ee
\be
2a_2 +2dF+2e(\frac{A^2}{m}+F^2)
=(2-m)D^2\,,
\ee
\be
2mc_2 B^2-e A^2=-mD^2\,.
\ee
Three of these equations
determine the three unknowns $A,B,D$ while the other three equations,
give three constraints between the seven parameters $a_{1,2},b_{1},
c_{1,2},d,e$.
In particular, $A$ and $B$ are given by
\be\label{3.48}
A^2=\frac{mD^2(e-2c_2)}{(e^2-4c_1 c_2)}\,, ~~~
B^2=\frac{D^2(2c_1-e)}{(e^2-4c_1 c_2)}\,,
\ee
while $D$ is given by
\be\label{3.49}
D^2=\frac{(e^2-4c_1c_2)a_1}{[(1+m)(e^2-4c_1c_2)+2e(2c_1-e)]}\,.
\ee
Further, the three constraints are
\be\label{3.50}
b_1^2=4a_1c_1\,,~~2a_1e+db_1=0\,,~~(2-m)mD^2=2a_2m+2eA^2+mdF\,.
\ee

\noindent{\bf m=1}

In this limiting case we have a bright-dark soliton solution given by
Eq. (\ref{3.38})
with $A$, $B$ and $D$ given by Eqs. (\ref{3.35}) and (\ref{3.36})
with $m=1$
while the three constraints are again given by Eq. (\ref{3.37}).

\noindent{\bf Special case of $e^2 = 4c_1 c_2$}

One can show that the solution (\ref{3.46}) exists even in case $e^2=
4c_1 c_2$. It turns out  such a solution
exists only if Eqs. (\ref{3.6}) and (\ref{3.47}) are satisfied.
In this case, $A$, $B$ and $D$ are given by
\be\label{3.52}
A^2= \frac{m[a_1+2(1-m)a_2]}{2e}\,,~~
B^2= \frac{[a_1-2(1+m)a_2]}{2e}\,,~~
D^2=2a_2\,,
\ee
while the constraints are
\be\label{3.53}
d=-b_1\,,~~b_1^2=4a_1c_1\,.
\ee

\noindent{\bf Interesting case of $b_2=c_2=0$}

Finally, let us discuss the physically interesting case
of $b_2=c_2=0$. 
In this case Eq. (\ref{3.46}) with $G=0$ is a solution to the
field Eq. (\ref{2.2}) provided
\be\label{3.54}
A^2=\frac{(d^2-4ea_2)}{2e^2}\,,~~B^2=\frac{(2c_1-e)(d^2-4ea_2)}{2me^3}\,,~~
D^2=\frac{(d^2-4ea_2)}{2me}\,,~~F=-\frac{d}{2e}\,,
\ee
and further if the following three constraints are satisfied
\be\label{3.55}
b_1^2=4a_1c_1\,,~~(d^2-4ea_2)[4c_1-(1-m)e]=2ma_1e^2\,,~~b_1d+2a_1e=0\,.
\ee

\noindent{\bf Energy:} Corresponding to the periodic solution
(\ref{3.46})
%\be
%\phi =F+A\cn[D(x+x_0),m]\,,~~\psi =B\cn[D(x+x_0),m]\,,
%\ee
the energy $\hat{E}$ and the constant $C$ are given by
\bea\label{3.56}
&&\hat{E}=\frac{2D}{3m}\bigg ([(2m-1)B^2+(1+m)A^2]E(m)
-(1-m)(A^2-B^2)K(m) \bigg )\,, \nonumber \\
&&C=F^2[a_1-b_1F+c_1F^2]-\frac{1}{2} A^2 D^2 +B^2[a_2+dF+eF^2+c_2 B^4]\,.
\eea

On using the expansion formulas for $E(m)$ and $K(m)$ around $m=1$ as
given by Eqs. (\ref{3.13}) and (\ref{3.14}),
for $m$ near one, the energy of the periodic solution can be rewritten
as the energy of the corresponding hyperbolic (bright-dark) soliton
solution [Eq. (\ref{3.38})]
%\be
%\phi =F+A\tanh[D(x+x_0)]\,,~~\psi =B\tanh[D(x+x_0)]\,,
%\ee
plus the interaction energy. We find
\be\label{3.57}
\hat{E}=E_{soliton}+E_{int}=D \left[\frac{2}{3}(2A^2+B^2)
+\frac{(1-m)}{6}(2A^2-5B^2)+(1-m)B^2\ln\left(\frac{4}{\sqrt{1-m}}\right)\right]\,.
\ee
Note that this solution exists when either (i) $2c_1\ge e \ge 2c_2$ and
$e^2 \ge 4c_1c_2$ or if (ii) $2c_2 \ge e \ge 2c_1$ and $4c_1c_2 \ge e^2$
or if (iii) $c_2=0, 2c_1 > e$.
The interaction energy
vanishes at exactly $m=1$, as it should.

\noindent{\bf Solution VI}

We now discuss two periodic solutions which at $m=1$ reduce to
dark-bright soliton solutions.
In particular, it is easily shown that
\be\label{3.58}
\phi =F+A\cn[D(x+x_0),m]\,,~~\psi =G+\sn[D(x+x_0),m]\,,
\ee
is an exact solution provided
\be\label{3.59}
G=0\,,~~b_2=0\,,~~ d+2eF=0\,,
\ee
 and the following six
 coupled equations are
satisfied
\be
2a_1 F-3b_1F^2+4c_1 F^3=0\,,
\ee
\be
2a_1 -6b_1F+12c_1 F^2 +2eB^2
=(2m-1)D^2\,,
\ee
\be
-3b_1A^2+12c_1 F A^2=0\,,
\ee
\be
2c_1 A^2-e B^2=-mD^2\,,
\ee
\be
2a_2 +2dF+2e(A^2+F^2)
=-(1+m)D^2\,,
\ee
\be
2c_2 B^2-e A^2=mD^2\,.
\ee
Three of these equations
determine the three unknowns $A,B,D$ while the other three equations,
give three constraints between the seven parameters $a_{1,2},b_{1},
c_{1,2},d,e$.
In particular, $A$ and $B$ are given by
\be\label{3.60}
A^2=\frac{mD^2(e-2c_2)}{(4c_1 c_2-e^2)}\,, ~~~
B^2=\frac{mD^2(2c_1-e)}{(4c_1 c_2-e^2)}\,,
\ee
while $D$ is given by
\be\label{3.61}
D^2=\frac{(4c_1c_2-e^2)a_1}{[(4c_1c_2-e^2)+4mc_1(e-2c_2)]}\,.
\ee
Further, the three constraints are
\be\label{3.62}
b_1^2=4a_1c_1\,,~~2a_1e+db_1=0\,,~~-(1+m)D^2=2a_2+2eA^2+dF\,.
\ee

\noindent{\bf m=1}

In this limiting case we have a dark-bright soliton solution given by
\be\label{3.63}
\phi =F+A\sech[D(x+x_0)]\,,~~\psi =B\tanh[D(x+x_0)]\,,
\ee
with $A$, $B$ and $D$ given by Eqs. (\ref{3.60}) and (\ref{3.61})
with $m=1$
while the three constraints are given by Eq. (\ref{3.62}).

\noindent{\bf Special case of $e^2 = 4c_1 c_2$}

One can show that the solution (\ref{3.58}) exists even in case $e^2=
4c_1 c_2$. It turns out  such a solution
exists only if Eqs. (\ref{3.6}) and (\ref{3.59}) are satisfied.
In this case, $A$, $B$ and $D$ are given by
\be\label{3.65}
A^2= \frac{(ma_1+2a_2)}{2em}\,,~~
B^2= \frac{[ma_1-2(2m-1)a_2]}{2em}\,,~~
D^2=-\frac{2a_2}{m}\,,
\ee
while the constraints are
\be\label{3.66}
d=-b_1\,,~~b_1^2=4a_1c_1\,.
\ee

Finally, let us discuss the physically interesting case
of $b_2=c_2=0$. In this case, a solution does not exist since whereas 
$A^2 >0$ requires that $e<0$ but that in turn makes $B^2 <0$, since from 
stability considerations, $c_1 >0$. 

\noindent{\bf Energy:} Corresponding to the periodic solution
(\ref{3.58})
%\be
%\phi =F+A\cn[D(x+x_0),m]\,,~~\psi =B\cn[D(x+x_0),m]\,,
%\ee
the energy $\hat{E}$ and the constant $C$ are given by
\bea\label{3.67}
&&\hat{E}=\frac{2D}{3m}\bigg ([(2m-1)A^2+(1+m)B^2]E(m)
-(1-m)(B^2-A^2)K(m) \bigg )\,, \nonumber \\
&&C=F^2[a_1-b_1F+c_1F^2]+\frac{1}{2}(1-m)A^2D^2+B^2[a_2+dF+eF^2+c_2B^2]\,.
\eea

On using the expansion formulas for $E(m)$ and $K(m)$ around $m=1$ as
given by Eqs. (\ref{3.13}) and (\ref{3.14}),
for $m$ near one, the energy of the periodic solution can be rewritten
as the energy of the corresponding hyperbolic (dark-bright) soliton
solution [Eq. (\ref{3.63})]
%\be
%\phi =F+A\tanh[D(x+x_0)]\,,~~\psi =B\tanh[D(x+x_0)]\,,
%\ee
plus the interaction energy. We find
\be\label{3.68}
\hat{E}=E_{soliton}+E_{int}=D \left[\frac{2}{3}(2B^2+A^2)
+\frac{(1-m)}{6}(2B^2-5A^2)+(1-m)A^2\ln\left(\frac{4}{\sqrt{1-m}}\right)\right]\,.
\ee
Note that this solution exists when either (i) $2c_1\ge e \ge 2c_2$ and
$4c_1c_2 \ge e^2$ or if (ii) $2c_2 \ge e \ge 2c_1$ and $e^2 \ge 4c_1c_2$.
The interaction energy
vanishes at exactly $m=1$, as it should.

\noindent{\bf Solution VII}

Another periodic solution which at $m=1$ reduces to the
dark-bright soliton solution is given by
\be\label{3.69}
\phi =F+A\dn[D(x+x_0),m]\,,~~\psi =G+B\sn[D(x+x_0),m]\,,
\ee
provided
\be\label{3.70}
G=0\,,~~b_2=0\,,~~ d+2eF=0\,,
\ee
 and the following six
 coupled equations are
satisfied
\be
2a_1 F-3b_1F^2+4c_1 F^3=0\,,
\ee
\be
2a_1 -6b_1F+12c_1 F^2 +\frac{2eB^2}{m}
=(2-m)D^2\,,
\ee
\be
-3b_1A^2+12c_1 F A^2=0\,,
\ee
\be
2mc_1 A^2-e B^2=-mD^2\,,
\ee
\be
2a_2 +2dF+2e(A^2+F^2)
=-(1+m)D^2\,,
\ee
\be
2c_2 B^2-me A^2=mD^2\,.
\ee
Three of these equations
determine the three unknowns $A,B,D$ while the other three equations,
give three constraints between the seven parameters $a_{1,2},b_{1},
c_{1,2},d,e$.
In particular, $A$ and $B$ are given by
\be\label{3.71}
A^2=\frac{D^2(e-2c_2)}{(4c_1 c_2-e^2)}\,, ~~~
B^2=\frac{mD^2(2c_1-e)}{(4c_1 c_2-e^2)}\,,
\ee
while $D$ is given by
\be\label{3.72}
D^2=\frac{(4c_1c_2-e^2)a_1}{[m(4c_1c_2-e^2)+4c_1(e-2c_2)]}\,.
\ee
Further, the three constraints are
\be\label{3.73}
b_1^2=4a_1c_1\,,~~2a_1e+db_1=0\,,~~-(1+m)D^2=2a_2+2eA^2+dF\,.
\ee

\noindent{\bf m=1}

In this limiting case we have a dark-bright soliton solution given by
Eq. (\ref{3.63}).

\noindent{\bf Special case of $e^2 = 4c_1 c_2$}

One can show that the solution (\ref{3.69}) exists even in case $e^2=
4c_1 c_2$. It turns out  such a solution
exists only if Eqs. (\ref{3.6}) and (\ref{3.70}) are satisfied.
In this case, $A$, $B$ and $D$ are given by
\be\label{3.75}
A^2= \frac{(a_1+2ma_2)}{2e}\,,~~
B^2= \frac{[a_1-2(2-m)a_2]}{2e}\,,~~
D^2=-2a_2\,,
\ee
while the constraints are
\be\label{3.76}
d=-b_1\,,~~b_1^2=4a_1c_1\,.
\ee

Finally, let us discuss the physically interesting case of $b_2=c_2=0$. 
As in the previous case, a solution does not exist since whereas 
$A^2 >0$ requires that $e<0$ but that in turn makes $B^2 <0$, since from 
stability considerations, $c_1 >0$. 

\noindent{\bf Energy:} Corresponding to the periodic solution
(\ref{3.69})
%\be
%\phi =F+A\cn[D(x+x_0),m]\,,~~\psi =B\cn[D(x+x_0),m]\,,
%\ee
the energy $\hat{E}$ and the constant $C$ are given by
\bea\label{3.77}
&&\hat{E}=\frac{2D}{3m}\bigg ([(2-m)mA^2+(1+m)B^2]E(m)
-(1-m)(B^2+2A^2)K(m) \bigg )\,, \nonumber \\
&&C=F^2[a_1-b_1F+c_1F^2]+\frac{1}{2}(1-m)A^2D^2+\frac{B^2}{m^2}
[ma_2+mdF+meF^2+c_2B^4]\,.
\eea

On using the expansion formulas for $E(m)$ and $K(m)$ around $m=1$ as
given by Eqs. (\ref{3.13}) and (\ref{3.14}),
for $m$ near one, the energy of the periodic solution can be rewritten
as the energy of the corresponding hyperbolic (dark-bright) soliton
solution [Eq. (\ref{3.63})]
%\be
%\phi =F+A\tanh[D(x+x_0)]\,,~~\psi =B\tanh[D(x+x_0)]\,,
%\ee
plus the interaction energy. We find
\be\label{3.78}
\hat{E}=E_{soliton}+E_{int}=D \left[\frac{2}{3}(2B^2+A^2)
+\frac{(1-m)}{6}(2B^2+3A^2)-(1-m)A^2\ln\left(\frac{4}{\sqrt{1-m}}\right)\right]\,.
\ee
Note that this solution exists when either (i) $2c_1\ge e \ge 2c_2$ and
$4c_1c_2 \ge e^2$ or if (ii) $2c_2 \ge e \ge 2c_1$ and $e^2 \ge 4c_1c_2$.
The interaction energy
vanishes at exactly $m=1$, as it should.
\vskip 2.5truecm 

{\bf Solution VIII}

We now present three solutions, which to the best of our knowledge were not
known before even in the uncoupled limit. Two of these solutions, in the
uncoupled limit are kink-type while one is a pulse type solution. Let us
discuss these solutions one by one.

One of these periodic solutions is given by
\be\label{3.79}
\phi =F+\frac{A\sn[D(x+x_0),m]}{1+B\dn[D(x+x_0),m]}\,,~~
\psi =G+\frac{H\sn[D(x+x_0),m]}{1+B\dn[D(x+x_0),m]}\,,~~
\ee
provided
\be\label{3.80}
B=1\,,
\ee
 and the following eight
 coupled equations are
satisfied
\be
2a_1 F-3b_1F^2+4c_1 F^3+dG^2+2eFG^2=0\,,
\ee
\be
2a_1A -6b_1FA+12c_1 F^2A +2dGH+4eFGH+2eAG^2
=-\frac{A(2-m)D^2}{2}\,,
\ee
\be
-3b_1A^2+12c_1 F A^2+dH^2+2eFH^2+4eAGH=0\,,
\ee
\be
8c_1 A^2+4e H^2=m^2D^2\,,
\ee
\be
2a_2 G-3b_2G^2+4c_2 G^3+2dFG+2eGF^2=0\,,
\ee
\be
2a_2H -6b_2GH+12c_2 G^2H +2dFH+2dAG+4eFGA+2eHF^2
=-\frac{H(2-m)D^2}{2}\,,
\ee
\be
-3b_2H^2+12c_2 G H^2+2dAH+2eGA^2+4eAFH=0\,,
\ee
\be
8c_2 H^2+4e A^2=m^2D^2\,. 
\ee

Five of these equations
determine the five unknowns $A,H,D,F,G$ while the other three equations,
give three constraints between the eight parameters $a_{1,2},b_{1,2},
c_{1,2},d,e$.
In particular, $A$ and $H$ are given by
\be\label{3.81}
A^2=\frac{m^2D^2(2c_2-e)}{4(4c_1 c_2-e^2)}\,, ~~~
H^2=\frac{m^2D^2(2c_1-e)}{4(4c_1 c_2-e^2)}\,.
\ee

It turns out that there is no solution to these field equations with
$F=G=0$ so long as $a_1>0$. However, there are solutions when either
$G=0, F \ne 0$ or $F=0, G \ne 0$ which we discuss one by one. 

\noindent{\bf G=0, F $\ne$ 0}

In this case $A,H$ are again given by Eq. (\ref{3.81}), $B=1$ while
$F$ and $D$ are given by
\bea\label{3.82}
&&F\equiv \phi_{m}=\frac{[3b_1-\sqrt{(9b_1^2-32a_1c_1)}]}{8c_1}\,, 
\nonumber \\
&&D^2=\frac{\sqrt{(9b_1^2-32a_1c_1)}[3b_1-\sqrt{(9b_1^2-32a_1c_1)}]}
{4(2-m)c_1}\,. 
\eea
Further there are three constraints given by Eq. (\ref{3.4}).

\noindent{\bf Special case of $e^2 = 4c_1 c_2$}

One can show that the solution (\ref{3.79}) with $G=0$, $a_1=b_1F$ 
exists even in case $e^2=
4c_1 c_2$ provided if Eq. (\ref{3.6}) is satisfied.
In this case, $A$ and $H$ can be  determined from the relations 
\be\label{3.86}
A^2+H^2=\frac{m^2D^2}{4e}\,,~~3b_2H=2(d+2eF)A\,,
\ee
$F,D$ are still given by (\ref{3.82}), while the two constraints
are given by Eq. (\ref{3.4a}).

\noindent{\bf Interesting case of $b_2=c_2=0$}

Finally, let us discuss the physically interesting case
of $b_2=c_2=0$. 
In this case Eq. (\ref{3.79}) with $G=0$ 
is a solution to the
field Eq. (\ref{2.2}) provided
\be\label{3.87}
A^2=\frac{m^2a_1}{2e(2-m)}\,,~~H^2=\frac{(e-2c_1)m^2a_1}{2e^2(2-m)}\,,
~~D^2=\frac{2a_1}{(2-m)}\,,~~F=\sqrt{\frac{a_1}{4c_1}}\,,
\ee
while the three constraints are  given by Eq. (\ref{3.8}).

Thus the solution with $G=0, F \ne 0$ exists when $2c_1\ge e, 2c_2 \ge e$ and
$4c_1c_2 \ge e^2$.

\noindent{\bf F=0, G $\ne$ 0}

Solution (\ref{3.79}) is a solution with $F=0, G \ne 0$ provided $d=0$,
$B=1$ and $e <0$, we therefore write $e=-|e|$. 
In this case $A,H$ are given by Eq. (\ref{3.81}) while
$G$ and $D$ are given by
\bea\label{3.82c}
&&G\equiv \psi_{m}=\frac{[3b_2-\sqrt{(9b_2^2-32a_2c_2)}]}{8c_2}\,, 
\nonumber \\
&&D^2=\frac{\sqrt{(9b_2^2-32a_2c_2)}[3b_2-\sqrt{(9b_2^2-32a_2c_2)}]}
{4(2-m)c_2}\,. 
\eea
Further there are three constraints given by
\be\label{3.82e}
3b_1A=-4|e|GH\,,~~3H^2(b_2-4c_2G)=-2|e|GA^2\,,~~
3(2c_2+|e|)b_2G=2(2a_1c_2+4a_2c_2+|e|a_2)\,.
\ee

For $F=0, G \ne 0$, no solution of the form (\ref{3.79}) exists
in case either $e^2=4c_1c_2$ or if $b_2=c_2=0$.

\noindent{\bf m=1}

In this limiting case we have a bright-bright soliton solution given by
\be\label{3.84}
\phi =F+A\tanh\left[\sqrt{\frac{a_1}{2}}(x+x_0)\right]\,,~~
\psi =H\tanh\left[\sqrt{\frac{a_1}{2}}(x+x_0)\right]\,,~~
\ee
with $A$ and $H$ given by Eq. (\ref{3.81}), $D^2=2a_1, B=1$ while the 
three constraints are again given by Eq. (\ref{3.4}).
\noindent{\bf Uncoupled Case}

For completeness, it may be worthwhile to write down the solution of the
uncoupled field Eq. (\ref{2.2}), i.e. when $d=e=0$. It is easily shown that
in this case the solution (\ref{3.79}) reduces to
\be\label{3.88}
\phi= \sqrt{\frac{a_1}{4c_1}}\left[1+\left(\frac{m}{2-m}\right)
\frac{\sn\left(\sqrt{\frac{2a_1}{2-m}}(x+x_0),m\right)}{\left[1+
\dn\left(\sqrt{\frac{2a_1}{2-m}}(x+x_0),m\right)\right]}\right]\,,
\ee
and one is at the transition temperature $T=T_c$ since the parameters 
$a_1,b_1,c_1$ satisfy $b_1^2=4a_1c_1$. At $m=1$, this reduces to the well 
known kink solution (\ref{3.10}).

\noindent{\bf Energy:} Corresponding to the periodic solution
(\ref{3.79}), while the energy is same but the value of $C$ is different
for solutions with either $G=0, F \ne 0$ or $F=0, G \ne 0$. For example, in 
case $G=0, F \ne 0$ 
%\be
%\phi =F+A\sn[D(x+x_0),m]\,,~~\psi =B\sn[D(x+x_0),m]\,,
%\ee
the energy $\hat{E}$ and the constant $C$ are given by
\bea\label{3.89}
&&\hat{E}=(A^2+H^2)D \int_{-2K}^{2K} \frac{\cn^2(y,m)}
{[1+\dn(y,m)]^2} dy\,, \nonumber \\
&&C=[a_1-b_1F+c_1F^2]F^2-\frac{1}{8}(A^2+H^2)D^2\,,
\eea
where $y=D(x+x_0)$. On the other hand, when $F=0, G \ne 0$, the constant 
$C$ is given by
\be
C=[a_2-b_2G+c_2G^2]G^2-\frac{1}{8}(A^2+H^2)D^2\,.
\ee

Unfortunately, we are unable to solve the integral (\ref{3.89}) 
analytically and hence
are unable to calculate the corresponding soliton interaction energy.
However, the energy of the corresponding hyperbolic (bright-bright) soliton
solution [Eq. (\ref{3.84})]
%\be
%\phi =F+A\tanh[D(x+x_0)]\,,~~\psi =B\tanh[D(x+x_0)]\,,
%\ee
is easily calculated. We find
\be\label{3.90}
E_{soliton}=\sqrt{\frac{16a_1}{18}} (A^2+H^2)D\,.
\ee
\vskip 2.0truecm 
{\bf Solution IX}

We now present a periodic solution, which at $m=1$
reduces to a constant but is nontrivial for arbitrary $m$.
This periodic solution is given by
\be\label{3.102}
\phi =F+\frac{A\cn[D(x+x_0),m]}{\sqrt{1-m}+B\dn[D(x+x_0),m]}\,,~~
\psi =G+\frac{H\cn[D(x+x_0),m]}{\sqrt{1-m}+B\dn[D(x+x_0),m]}\,,~~
\ee
provided Eq. (\ref{3.80}) holds (i.e. $B=1$)
 and the following eight coupled equations are
satisfied
\be
2a_1 F-3b_1F^2+4c_1 F^3+dG^2+2eFG^2=0\,,
\ee
\be
2a_1A -6b_1FA+12c_1 F^2A +2dGH+4eFGH+2eAG^2
=-\frac{A(2-m)D^2}{2}\,,
\ee
\be
-3b_1A^2+12c_1 F A^2+dH^2+2eFH^2+4eAGH=0\,,
\ee
\be
8c_1 A^2+4e H^2=m^2D^2\,,
\ee
\be
2a_2 G-3b_2G^2+4c_2 G^3+2dFG+2eGF^2=0\,,
\ee
\be
2a_2H -6b_2GH+12c_2 G^2H +2dFH+2dAG+4eFGA+2eHF^2
=-\frac{H(2-m)D^2}{2}\,,
\ee
\be
-3b_2H^2+12c_2 G H^2+2dAH+2eGA^2+4eAFH=0\,,
\ee
\be
8c_2 H^2+4e A^2=m^2D^2\,.  
\ee

It is rather remarkable that the eight field equations that we obtain
here are identical to those obtained for the solution VIII. As a result,
all the solutions obtained in that case continue to be true even in this 
case. In particular, as in the previous case, there does not exist a solution
to Eq. (\ref{3.102}) in case $F=G=0$. On the other hand, when $G=0, F \ne 0$
or when $F=0, G \ne 0$, the solutions exist 
with the parameters as given for solution VIII. 

\noindent{\bf m=1}

In this limit, (unlike the solution VIII) the solution (\ref{3.102}) 
with $G=0, F \ne 0$ goes over to a constant solution, given by
\be\label{3.107}
\phi=F+A\,,~~\psi=H\,,
\ee
where $A,H$ are given by Eq. (\ref{3.81}) while $F=\phi_{m}$ with 
$\phi_{m}$ given by Eq. (\ref{2.4a}). 
Similar conclusions are also obviously valid for the solution with 
$F=0, G \ne 0$.

\noindent{\bf Uncoupled Case}

For completeness, it may be worthwhile to write down the solution of the
uncoupled field Eq. (\ref{2.2}), i.e. when $d=e=0$. It is easily shown that
in this case the solution (\ref{3.102}) reduces to
\be\label{3.111}
\phi= \sqrt{\frac{a_1}{4c_1}}\left(1\pm \frac{m}{\sqrt{(2-m)}}
\frac{\cn\left[\sqrt{\frac{2a_1}{2-m}}(x+x_0),m\right]}{\left(\sqrt{1-m}+
\dn\left[\sqrt{\frac{2a_1}{2-m}}(x+x_0),m\right]\right)}\right)\,,
\ee
and one is at the transition temperature $T=T_c$ since the parameters 
$a_1,b_1,c_1$ satisfy $b_1^2=4a_1c_1$. At $m=1$, $\phi$ is a constant, 
which is either 0 or $\phi_c=\sqrt{\frac{a_1}{c_1}}$, i.e. at $m=1$, the 
field $\phi$ stays at one of the degenerate minima of the potential.

\noindent{\bf Energy:} Corresponding to the periodic solution
(\ref{3.102}), while the energy is same, the value of $C$ differs
depending on if $G=0, F \ne 0$ or if $F=0, G \ne 0$. For example, if
$G=0, F \ne 0$,  
%\be
%\phi =F+A\sn[D(x+x_0),m]\,,~~\psi =B\sn[D(x+x_0),m]\,,
%\ee
the energy $\hat{E}$ and the constant $C$ are given by
\bea\label{3.112}
&&\hat{E}=(1-m)(A^2+H^2)D \int_{-2K}^{2K} \frac{\sn^2(y,m)}
{[\sqrt{1-m}+\dn(y,m)]^2} dy\,, \nonumber \\
&&C=[a_1-b_1F+c_1F^2]F^2-\frac{1}{8}(A^2+H^2)D^2\,,
\eea
where $y=D(x+x_0)$. ON the other hand, if $F=0, G \ne 0$, then $C$ is given by
\be
C=[a_2-b_2G+c_2G^2]G^2-\frac{1}{8}(A^2+H^2)D^2\,,
\ee

Unfortunately, we are unable to solve the integral (\ref{3.112}) 
analytically and hence
are unable to calculate the corresponding soliton interaction energy.
However, the energy of the corresponding solution [Eq. (\ref{3.107})] is 
of course zero, being a constant solution.

{\bf Solution X}

We now discuss a periodic solution which in the uncoupled limit and even at
$m=1$ is distinct from the well known kink solution (\ref{3.10}).
This periodic solution is given by
\be\label{3.114}
\phi =F+\frac{A\sn[D(x+x_0),m]}{1+B\sn[D(x+x_0),m]}\,,~~
\psi =G+\frac{H\sn[D(x+x_0),m]}{1+B\sn[D(x+x_0),m]}\,,~~
\ee
provided
 the following eight
 coupled equations are
satisfied
\be
2a_1 F-3b_1F^2+4c_1 F^3+dG^2+2eFG^2=-2BAD^2\,,
\ee
\be
2a_1A -6b_1FA+12c_1 F^2A +2dGH+4eFGH+2eAG^2
=A[6B^2-(1+m)]D^2\,,
\ee
\be
-3b_1A^2+12c_1 F A^2+dH^2+2eFH^2+4eAGH=3AB[(1+m)-2B^2]D^2\,,
\ee
\be
2c_1 A^2+e H^2=(1-B^2)(m-B^2)D^2\,,
\ee
\be
2a_2 G-3b_2G^2+4c_2 G^3+2dFG+2eGF^2=-2BHD^2\,,
\ee
\be
2a_2H -6b_2GH+12c_2 G^2H +2dFH+2dAG+4eFGA+2eHF^2
=H[6B^2-(1+m)]D^2\,,
\ee
\be
-3b_2H^2+12c_2 G H^2+2dAH+2eGA^2+4eAFH=3HB[(1+m)-2B^2]D^2\,,
\ee
\be
2c_2 H^2+e A^2=(1-B^2)(m-B^2)D^2\,.  
\ee

Six of these equations determine the six unknowns $A,B,H,D,F,G$ while the 
other two equations, give two constraints between the eight parameters 
$a_{1,2},b_{1,2}, c_{1,2},d,e$.  In particular, $A$ and $H$ are given by
\be\label{3.115}
A^2=\frac{(1-B^2)(m-B^2)D^2(2c_2-e)}{(4c_1 c_2-e^2)}\,, ~~~
H^2=\frac{(1-B^2)(m-B^2)D^2(2c_1-e)}{(4c_1 c_2-e^2)}\,.
\ee
Unfortunately, it is not very easy to identify a case when the analysis is
somewhat simpler. One possible case is when $F=0$. In that case, it is
easily seen that five of the above equations determine the five unknowns
$A,B,G,H,D$ while the remaining three equations give three constraints.
Further, solutions also exist when $2c_1=2c_2=e$ as well as when $b_2=c_2=0$.
However, because of lack of simplicity of these solutions, we do not write 
them here explicitly.

\noindent{\bf Uncoupled case at $m=1$}

There is one case when the analysis is simple and an explicit solution
can be written down. In particular, in the uncoupled case, at $m=1$
it is easily shown that
\be\label{3.116}
\phi= \sqrt{\frac{a_1}{4c_1}}(1+B)
\left(\frac{1+\tanh\left[\sqrt{\frac{a_1}{2}}(x+x_0)\right]}{1
+B\tanh\left[\sqrt{\frac{a_1}{2}}(x+x_0)\right]} \right)\,,
\ee
is an exact solution to the above equations.
Note that
in this case
\be\label{3.117}
b_1^2=4a_1c_1\,,~~A=F(1-B)\,,~~Fb_1=a_1(1+B)\,,~~2D^2=a_1\,,
\ee
so that one is again at $T=T_c$. Thus one has generalized the well known 
kink solution at $T=T_c$ by obtaining a one parameter family of kink solutions,
characterized by the parameter $B$. In the limit $B=0$ this solution goes 
over to the well known kink solution but is different otherwise.
In particular, while as $x \rightarrow \pm \infty$,
both solutions (\ref{3.10}) and (\ref{3.116}) go to $\phi_c \equiv
\sqrt{\frac{a_1}{c_1}}$ and 0 respectively, the difference is in how they
behave at $x=0$. In particular, while the kink solution (\ref{3.10}) goes
to $\phi_{max} \equiv \sqrt{\frac{a_1}{2c_1}}$, the new kink solution
(\ref{3.117}) goes to $\phi=(1+B)\sqrt{\frac{a_1}{2c_1}}$. In other words,
with the new solution, the center of the kink can be anywhere between $0$
and $\phi_c$, while for the conventional kink solution (\ref{3.10}), the
center of the kink is always at $\phi_{max}$.
However, the energy of all the kink solutions is identical, i.e.
$E_{kink}=\frac{a_1^{3/2}}{3\sqrt{2}c_1}$ and is independent of $B$.

\noindent{\bf Energy:} Corresponding to the periodic solution
(\ref{3.114}) 
%\be
%\phi =F+A\sn[D(x+x_0),m]\,,~~\psi =B\sn[D(x+x_0),m]\,,
%\ee
the energy $\hat{E}$ and the constant $C$ are given by
\bea\label{3.118}
&&\hat{E}=(A^2+H^2)D \int_{-2K}^{2K} \frac{\cn^2(y,m)\dn^2(y,m)}
{[1+B\sn(y,m)]^4} dy\,, \nonumber \\
&&C=\left(F+\frac{A}{B}\right)^2\left[a_1-b_1\left(F+\frac{A}{B}\right) 
+c_1\left(F+\frac{A}{B}\right)^2+e\left(G+\frac{H}{B}\right)^2\right] \nonumber \\
&&+\left(G+\frac{H}{B}\right)^2\left[a_2-b_2\left(G+\frac{H}{B}\right) 
+c_2\left(G+\frac{H}{B}\right)^2 +d\left(F+\frac{A}{B}\right)\right]\,, 
\eea
where $y=D(x+x_0)$.
Unfortunately, we are unable to solve this integral analytically and hence
are unable to calculate the corresponding soliton interaction energy.
However, the energy of the corresponding hyperbolic (bright-bright) soliton
solution [Eq. (195)]    
\be\label{3.119}
\phi =F+\frac{A\tanh[D(x+x_0)]}{1+B\tanh[D(x+x_0)]}\,,~~
\psi =G+\frac{H\tanh[D(x+x_0)]}{1+B\tanh[D(x+x_0)]}\,,~~
\ee
is easily computed. We find that
\be\label{3.120}
E_{soliton}=\frac{4(A^2+H^2)D}{3(1-B^2)^2}\,.
\ee

Note that this solution exists when $2c_1\ge e, 2c_2 \ge e$ and
$4c_1c_2 \ge e^2$.

\section{Solutions when $b_1^2 \ne 4a_1c_1$ (in the Analogous Uncoupled 
Case): Solution XI}

So far, we have considered ten solutions, all of which, in the uncoupled
limit correspond to $b_1^2=4a_1c_1$, i.e. $T=T_c$. Now we will discuss
a case when the uncoupled limit at $m=1$ corresponds to $T \ne T_c$, i.e.
one is either above or below the transition temperature $T_c$ and as a 
consequence one has a pulse lattice solution.  In particular, it is easily 
shown that
\be\label{4.1}
\phi =F+\frac{A\dn[D(x+x_0),m]}{1+B\dn[D(x+x_0),m]}\,,~~
\psi =G+\frac{H\dn[D(x+x_0),m]}{1+B\dn[D(x+x_0),m]}\,,
\ee
is an exact solution to the field Eq. (\ref{2.2}) provided the following 
eight coupled equations are satisfied
\be
2a_1 F-3b_1F^2+4c_1 F^3+dG^2+2eFG^2=2(1-m)BAD^2\,,
\ee
\be
2a_1A -6b_1FA+12c_1 F^2A +2dGH+4eFGH+2eAG^2
=A[2-m-6(1-m)B^2]D^2\,,
\ee
\be
-3b_1A^2+12c_1 F A^2+dH^2+2eFH^2+4eAGH=-3AB[(2-m)-2(1-m)B^2]D^2\,,
\ee
\be
2c_1 A^2+e H^2=(B^2-1)[1-(1-m)B^2]D^2\,,
\ee
\be
2a_2 G-3b_2G^2+4c_2 G^3+2dFG+2eGF^2=2(1-m)BHD^2\,,
\ee
\be
2a_2H -6b_2GH+12c_2 G^2H +2dFH+2dAG+4eFGA+2eHF^2
=H[(2-m)-6(1-m)B^2]D^2\,,
\ee
\be
-3b_2H^2+12c_2 G H^2+2dAH+2eGA^2+4eAFH=-3HB[(2-m)-2(1-m)B^2]D^2\,,
\ee
\be
2c_2 H^2+e A^2=(B^2-1)[1-(1-m)B^2]D^2\,. 
\ee

Six of these equations determine the six unknowns $A,B,H,D,F,G$ while the 
other two equations, give two constraints between the eight parameters 
$a_{1,2},b_{1,2}, c_{1,2},d,e$.  In particular, $A$ and $H$ are given by
\be\label{4.2}
A^2=\frac{(B^2-1)[1-(1-m)B^2]D^2(2c_2-e)}{(4c_1 c_2-e^2)}\,, ~~~
H^2=\frac{(B^2-1)[1-(1-m)B^2]D^2(2c_1-e)}{(4c_1 c_2-e^2)}\,,
\ee
from where it follows that $B^2$ must satisfy the bound
\be\label{4.2a}
1 < B^2 < \frac{1}{1-m}\,.
\ee
So far, we have not been able to obtain explicit solutions to these equations
in an elegant form except at $m=1$ which we now present.

{\bf m=1}

In case $m=1$, the solution (\ref{4.1}) goes over to the pulse solution
\be\label{4.1a}
\phi =F+\frac{A\sech[D(x+x_0),m]}{1+B\sech[D(x+x_0),m]}\,,~~
\psi =G+\frac{H\sech[D(x+x_0),m]}{1+B\sech[D(x+x_0),m]}\,,
\ee
where $A,H$ are now given by
\be\label{4.3}
A^2=\frac{(B^2-1)D^2(2c_2-e)}{(4c_1 c_2-e^2)}\,, ~~~
H^2=\frac{(B^2-1)D^2(2c_1-e)}{(4c_1 c_2-e^2)}\,.
\ee
Several solutions are possible in this case, which we discuss one by one.

{\bf (i) $F=G=0,m=1$}

In this case one finds that while $A,H$ are given by Eq. (\ref{4.3}), the 
inverse characteristic length $D$ is given by
\be\label{4.4}
D^2=2a_1\,,~~a_1=a_2\,,
\ee
and $B$ is determined from the relation
\be\label{4.5}
dH^2+6ABa_1=3b_1A^2\,.
\ee
Further, the parameters satisfy the constraint
\be\label{4.6}
(2d+3b_1)(2c_2-e)-d(2c_1-e)=3b_2\sqrt{(2c_1-e)(2c_2-e)}\,.
\ee

{\bf Interesting case of $b_2=c_2=0$}

In this physically interesting case, one finds that
\be\label{4.7}
A^2=\frac{(B^2-1)D^2}{e}\,, ~~~
H^2=\frac{(B^2-1)D^2(2c_1-e)}{e^2}\,,
\ee
while Eq. (\ref{4.4}) is still valid. However, $B$ is now given by a
simpler expression
\be\label{4.8}
B^2=\frac{2d^2}{(2d^2-9ea_1)}\,,
\ee
and the constraint has a simpler form
\be\label{4.9}
d(e+2c_1)+3b_1e=0\,.
\ee

{\bf Special case of $4c_1c_2=e^2$}

The above solution continues to exist even in the case $4c_1c_2=e^2$
provided $2c_1=2c_2=e$. Further, while Eq. (\ref{4.4}) is still valid,
 $A,H$ are no more given by
Eq. (\ref{4.3}), but they satisfy the constraint relations
\be\label{4.10}
e(A^2+H^2)=(B^2-1)D^2\,,~~(2d+3b_1)A^2=dH^2+3b_2HA\,,~~2(dA+3Ba_1)=
3b_2H\,,
\ee
from where one can determine $A,H$ and $B$.

{\bf (ii) $G=0, F \ne 0, m=1$}

In this case one finds that while $A,H$ are given by Eq. (\ref{4.3}), $D$ 
and $F$ are now given by
\be\label{4.11}
D^2=3b_1F-4a_1\,,~~F\equiv \phi_{c}
=\frac{3b_1+\sqrt{(9b_1^2-32a_1c_1)}}{8c_1}\,.
\ee
From the remaining three equations, one can determine $B$ and further one
has two constraints between the various parameters. This solution is also
valid in case $b_2,c_2=0$ or when $2c_1=2c_2=e$ with appropriate constraints
which can be easily worked out.

{\bf (iii) $F=0, G \ne 0, m=1$}

In this case one finds that the solution exists only if $d=0$. Further,
 $A,H$ are still given by Eq. (\ref{4.3}), while $D$ 
and $G$ are now given by
\be\label{4.11c}
D^2=3b_2G-4a_2\,,~~G\equiv \psi_{c}
=\frac{3b_2+\sqrt{(9b_2^2-32a_2c_2)}}{8c_2}\,.
\ee
From the remaining three equations, one can determine $B$ and further one
has two constraints between the various parameters. This solution is 
however not valid either in case $b_2,c_2=0$ or when $2c_1=2c_2=e$.

{\bf Uncoupled case with $m=1$}

Finally, for completeness, we write down the solution in the uncoupled case
when $m=1$. In this case, it is not difficult to show that one
of the solution is given by
\be\label{4.12}
\phi =\frac{A\sech[\sqrt{2a_1}(x+x_0)]}{1+B\sech[\sqrt{2a_1}(x+x_0)]}\,,
\ee
where
\be\label{4.13}
B=\frac{b_1}{\sqrt{b_1^2-4a_1c_1}}\,,~~
A=\frac{2a_1}{\sqrt{b_1^2-4a_1c_1}}\,.
\ee
The solution (\ref{4.12}) can also be written in the form \cite{konwent}
\be\label{4.14}
\phi=\frac{2a_1}{b_1+\sqrt{(b_1^2-4a_1c_1)}\,cosh[\sqrt{2a_1}(x+x_o)]}\,.
\ee
Note that this solution exists provided $b_1^2>4a_1c_1$, i.e. below the 
transition temperature $T<T_c$.  As pointed out at several places in the 
literature, this solution is a pulse solution around the local minimum 
$\phi=0$.

The uncoupled equations also have another solution given by
\be\label{4.15}
\phi =F+\frac{A\sech[D(x+x_0),m]}{1+B\sech[D(x+x_0),m]}\,,
\ee
where
\bea\label{4.16}
&&F\equiv \phi_c 
=\frac{3b_1+\sqrt{(9b_1^2-32a_1c_1)}}{8c_1}\,,~~D^2=3b_1F-4a_1\,,
\nonumber \\
&&B=\frac{4c_1F-b_1}{\sqrt{(b_1^2-2b_1c_1F)}}\,,~~
A=-\frac{3b_1F-4a_1}{\sqrt{(b_1^2-2b_1c_1F)}}\,.
\eea
Note that this solution is valid only if $b_1^2<4a_1c_1$, i.e. above the 
transition temperature $T>T_c$.  This follows from the fact that 
$b_1^2>2b_1c_1F$ so that $A,B$ are real.  As pointed out at several places 
in the literature, this solution is a pulse solution around the local 
minimum $\phi=\phi_c \equiv F$ where $\phi_c$ is as given by Eq. (\ref{2.3a}).

Summarizing, we have obtained eleven solutions to the coupled Eq. (\ref{2.2})
in case $a_1>0$ out of which ten solutions, for the corresponding uncoupled 
limit, are at $T=T_c$ (i.e. $b_1^2=4a_1c_1$) while one solution is for 
$T > T_C$ or $T<T_c$ depending on the value of the parameters. 

\section{Solutions when $a_1 <0$}

Before concluding this paper, we consider a few solutions of the coupled Eq. 
(\ref{2.2}) in case $a_1<0$. As pointed out in Sec. II, the uncoupled model 
with $a_1<0$ is also a model for first order transition with the temperature 
now being controlled by the parameter $b_1$. We will see that most of the 
solutions which we have obtained in the case $a_1>0$, are also valid when 
$a_1<0$ but with suitable modifications of parameters.

{\bf Solution I with $F=G=0$}

In Sec. III we saw that with $a_1>0$, solution I as given by Eq. (\ref{3.1})
is not possible in case $F=G=0$. However, it turns out that if $a_1<0$ then
such a solution is indeed possible. $A$ and $B$ are still given by 
Eq. (\ref{3.2}) while $D$ is now given by
\be\label{5.1}
D^2=\frac{2|a_1|}{(1+m)}\,,
\ee
while the three constraints are
\be\label{5.2}
a_1=a_2\,,~~27b_2^2b_1=4d^3\,,~~\frac{3b_1}{d}=\frac{(2c_1-e)}{(2c_2-e)}\,.
\ee

{\bf Solution I with $G=0, F \ne 0$}

Similarly, by looking at Eqs. (\ref{3.3}) and (\ref{3.4}) 
it is easy to convince oneself that  solution (\ref{3.1}) with $G=0$ and 
$F=\phi_{c1}$ or $F=\phi_{c2}$ exists in case $a_1<0$ and the parameters
satisfy essentially similar relations  to Eqs. (\ref{3.2}) to (\ref{3.4}) with
appropriate modifications. Further, such  solutions continue to exist in
the special case of $e^2=4c_1c_2$. However, such  solutions do not exist
in the physically interesting limit of $b_2=c_2=0$. Note that $\phi_{c1}$, 
$\phi_{c2}$ are as defined by Eqs. (\ref{2.5a}) and (\ref{2.6}).
 
{\bf Solution I with $F=0, G \ne 0$}

Similarly, by looking at Eqs. (\ref{3.3c}) and (\ref{3.4c}) 
it is easy to convince oneself that  solution (\ref{3.1}) with $F=0$ and 
$G=\psi_{m}$ exists in case $a_1<0$ and the parameters
satisfy essentially similar relations  to Eqs. (\ref{3.2c}) to (\ref{3.4c}) 
with appropriate modifications. But, such a solution does not exist either
when $e^2=4c_1c_2$ or when $b_2=c_2=0$. Note that $\psi_{m},\psi_{c}$ are 
essentially same as $\phi_m,\phi_c$ but with the replacement of $a_1,b_1,c_1$
by $a_2,b_2,c_2$.

{\bf Solution II with $F=G=0$}

In Sec. III we had seen that solution (\ref{3.16}) with $F=G=0$ and $m<1/2$
is not possible in case $a_1 >0$. It is however easy to see that if $a_1<0$,
then such a solution is clearly possible (with $m<1/2$) provided relations 
(\ref{3.17}) to (\ref{3.19}) are satisfied.  However, like other solutions 
of type II (and also III) with $a_1>0$, no solutions of type II with $a_1<0$ 
are possible in case either $e^2=4c_1c_2$ or if $b_2=c_2=0$.

{\bf Solution II with $G=0, F \ne 0$}

Similarly, by looking at Eqs. (\ref{3.20}) and (\ref{3.21}) 
it is easy to convince oneself that  solution (\ref{3.16}) with $m>1/2$,
$G=0$ and 
$F=\phi_{c1}$ or $F=\phi_{c2}$ exists in case $a_1<0$ and the parameters
satisfy essentially similar relations  to Eqs. 
(\ref{3.17}), (\ref{3.20}) and  (\ref{3.21}) with
appropriate modifications. 

{\bf Solution II with $F=0, G \ne 0$}

Similarly, by looking at Eqs. (\ref{3.20c}) and (\ref{3.21c}) 
it is easy to convince oneself that  solution (\ref{3.16}) with $m<1/2$,
$F=d=0$ and 
$G=\psi_{m}$ exists in case $a_1<0$ and the parameters
satisfy essentially similar relations  to Eqs. 
(\ref{3.20c}), (\ref{3.21c}) and  (\ref{3.21e}) with
appropriate modifications. 

{\bf Solution III with $G=0, F \ne 0$}

Looking at Eqs. (\ref{3.29}) and (\ref{3.30}) 
it is easy to convince oneself that  solution (\ref{3.25}) with $G=0$ and 
$F=\phi_{c1}$ or $F=\phi_{c2}$ exists for any $m$ ($0 \le m \le 1$),
in case $a_1<0$ and the parameters
satisfy essentially similar relations  to Eqs. 
(\ref{3.26}), (\ref{3.29}) and  (\ref{3.30}) with
appropriate modifications. 

{\bf Solution VIII with $F=G=0$}

In Sec. III we observed that with $a_1>0$, solution VIII as given by 
Eq. (\ref{3.79}) is not possible in case $F=G=0$. However, it turns out 
that if $a_1<0$ then such a solution is indeed possible. $A$ and $H$ are 
still given by Eq. (\ref{3.81}) while the inverse characteristic length 
$D$ is now given by
\be\label{5.3}
D^2=\frac{4|a_1|}{(2-m)}\,,
\ee
while the three constraints are
\be\label{5.4}
a_1=a_2\,,~~27b_2^2b_1=4d^3\,,~~\frac{3b_1}{d}=\frac{(2c_1-e)}{(2c_2-e)}\,.
\ee
 
{\bf Solution VIII with $F=0, G \ne 0$}

While no solutions are possible for $G=0, F \ne 0$ in case $a_1<0$, it
turns out that solutions are indeed possible in case $F=0, G \ne 0, a_1<0$
provided $d=0$ and $B=1$. 
In this case $A,H$ are given by Eq. (\ref{3.81}) while
$G$ and $D$ are given by Eq. (\ref{3.82c}). 
Further there are three constraints given by
\be\label{3.82g}
3b_1A=4eGH\,,~~3H^2(b_2-4c_2G)=2eGA^2\,,~~
4(|a_1-eG^2)=(2-m)D^2\,.
\ee

{\bf Solution IX with $F=G=0$}

In Sec. III we also observed that with $a_1>0$, solution IX as given by 
Eq. (\ref{3.102}) is not possible in case $F=G=0$. However, it turns out 
that if $a_1<0$ then such a solution is indeed possible. $A$ and $H$ are 
still given by Eq. (\ref{3.81}) while $D$ and the three constraints are 
again given by Eqs. (\ref{5.3}) and (\ref{5.4}), respectively.

{\bf Solution IX with $F=0, G \ne 0$}

Since solutions VIII and IX satisfy same field equations, it is then clear
that a solution with $F=0, G \ne 0, a_1<0$ will also exist in this case.

{\bf Solution XI}

In the last section we obtained a pulse-like periodic solution in case
$a_1>0$. It is amusing to note that the same solution continues to be valid 
even when $a_1<0$. In particular, note that in the uncoupled limit and
at $m=1$ the solutions for $a_1<0$ are again pulse solutions around the local
minimum. Specifically, if $b_1 > (<)$ 0, then the pulse solution is around
$\phi_{c2}$ $(\phi_{c1})$. In both cases, the solution is given by 
Eq. (\ref{4.15}) but with F now being either $\phi_{c1}$ or $\phi_{c2}$
where $\phi_{c1}$, $\phi_{c2}$ are as given by Eqs. (\ref{2.5a}) and 
(\ref{2.6}), respectively, and the values of other parameters like $D,B,A,H$ 
are again given by Eq. (\ref{4.16}) but with appropriate modification.

\section{Conclusion}

We have systematically provided an exhaustive set of exact periodic
domain wall solutions for a coupled asymmetric double well model (with
and) without an external field.  We found ten solutions at the transition 
temperature ($T=T_c$) and one solution each for $T>T_c$ and $T<T_c$. For 
the (physically interesting) special case when $b_2=c_2=0$ there is no 
nonlinearity in the $\psi$ field. However, due to the coupling with the 
$\phi$ field which has explicit nonlinearity, an effective nonlinearity 
is induced in the $\psi$ field thus enabling soliton-like solutions.  
When we set the coupling parameters $d=e=0$ the soliton solutions cease 
to exist in the $\psi$ field, as expected (for $b_2=c_2=0$).  We emphasize 
that the solutions found here are quite different from the ones we found 
for the first order transition in the coupled $\phi^6$ model \cite{cphi6} 
or the second order transition in coupled symmetric double well ($\phi^4$) 
model \cite{cphi4}.  

It would be instructive to explore whether the different solutions reported 
here in Sec. III are completely disjoint or if there are any possible
bifurcations linking them via, for instance, analytical continuation.
We have not tried to carry out an explicit stability analysis of various
periodic solutions.  However, the energy calculations and interaction
energy between solitons (for $m\sim 1$) could provide useful insight in
this direction.  Unlike the coupled $\phi^4$ and $\phi^6$ cases 
\cite{cphi4,cphi6}, we have not yet succeeded in finding a solvable 
discrete analog of an asymmetric double well model.

Our results are relevant for a wide range of physical situations
including structural transformations (with coupling of strain and
shuffle modes) \cite{jacobs,morse}, liquid crystals \cite{liquid}, 
hydrogen bonded chains \cite{xu1,xu2,xu3,cheng} and field theoretic 
contexts \cite{field}.  For structural phase transformations the 
solutions provide novel domain wall arrays, e.g. periodic antiphase 
boundaries and twin boundaries. Similarly, for hydrogen bonded chains 
these solutions represent new periodic nonlinear excitations.    

\section{Acknowledgment}
A.K. acknowledges the hospitality of the Center for Nonlinear Studies 
and the Theoretical Division at LANL.  This work was supported in part 
by the U.S.  Department of Energy.

\end{document}